\documentclass[aps,twocolumn,amsmath,amssymb,showpacs,prb]{revtex4}
\usepackage{epsf}
\newcommand{\ki}{\hat{\bf p}}
\newcommand{\ko}{\hat{\underline{\bf p}}}
\newcommand{\vR}{{\bf v}_f}
\begin{document}
\title{Transfer-matrix description of heterostructures involving
  superconductors and ferromagnets} 
\author{J.~Kopu$^1$, 
M.~Eschrig$^1$, J.~C.~Cuevas$^1$, and M.~Fogelstr\"om$^2$} 
\affiliation{$^1$Institut f\"{u}r Theoretische
Festk\"{o}rperphysik, Universit\"{a}t Karlsruhe, D-76128 Karlsruhe,
Germany \\
$^2$Applied Quantum Physics and Complex Systems, MC2, Chalmers,
S-41296 G\"oteborg, Sweden
} 
\date{\today} 
\begin{abstract}
  Based on the technique of quasiclassical Green's functions, we
  construct a theoretical framework for describing heterostructures
  consisting of superconductors and/or spin-polarized materials. The
  necessary boundary conditions at the interfaces separating different
  metals are formulated in terms of hopping amplitudes in a $t$-matrix
  approximation. The theory is applicable for an interface with
  arbitrary transmission and exhibiting scattering with arbitrary spin
  dependence. Also, it can be used in describing both ballistic and
  diffusive systems. We establish the connection between the standard
  scattering-matrix approach and the existing boundary conditions, and
  demonstrate the advantages offered by the $t$-matrix description.
\end{abstract}
\maketitle

\section{Introduction}

Low-energy (below the superconducting energy gap) electron transport
through contacts between a superconductor and a normal metal can be
understood in terms of Andreev reflection.\cite{Andreev} In this
process, an incident electron from the normal side can enter the
superconductor by pairing with another electron with the opposite
spin, leaving a reflected hole in the normal metal. The phase-coherent
nature of this process results in superconducting correlations being
induced in the normal-metal side, referred to as the proximity effect.
The important feature of Andreev reflection is that, with singlet
superconductors, it involves both spin bands in the normal metal.
Therefore, the above simple picture has to be modified when the normal
metal is replaced by a ferromagnet with two different Fermi surfaces
for the two spins, resulting in new and interesting physical
phenomena. In recent years, interplay between superconductivity and
ferromagnetism has attracted considerable theoretical
\cite{Falko,Bergeret,Kadigrobov,Huertas} and experimental
\cite{Giroud,Petrashov,Aumentado,Kontos} attention -- both out of
fundamental scientific interest and in view of the possibility of
novel applications and devices. One important consequence of the spin
splitting between the two bands in the ferromagnet is that the phase
coherence between the particle-hole pair in the clean (dirty) limit is
destroyed over a characteristic distance of $v_f/h$ ($\sqrt{D/h}$),
where $v_f$ is the Fermi velocity, $D$ the diffusion constant, and $h$
an effective exchange energy which describes the spin splitting.
Since this distance is typically very short (of the order of atomic
distances), the superconducting correlations induced to the
ferromagnetic material are expected to be confined to the immediate
vicinity of the separating interface. This raises the question
whether, for strong ferromagnets, a mechanism of a different type
takes over and dominates the physics of superconductor/ferromagnet
(S/F) contacts. One such mechanism, recently under active
investigation, is the inducement of spin-triplet correlations: namely,
the exchange field only affects correlations of singlet type, {\it
  i.e} between particles and holes in opposite spin bands.  In fact,
equal-spin triplet correlations are expected to be created by
proximity to a ferromagnet due to the breaking of spin-rotational
symmetry. The desire to formulate a theory capable of understanding
the detailed nature and the conditions for the formation of these
correlations and the corresponding anomalous proximity effect has
given the initial motivation for this work.

Problems related to superconducting proximity effect with
spin-dependent interfacial scattering are of spatially inhomogeneous
nature and, as such, they can only be studied with specialized
theoretical tools. Such a tool is provided by the quasiclassical
theory of superconductivity.\cite{Eilenberger,Larkin} This theory is
applicable for weakly perturbed superconductors (characteristic length
scale of perturbations much larger than Fermi wave length and
characteristic frequencies much less than Fermi energy) and can be
used in both equilibrium and nonequilibrium situations. It describes
quasiparticles with momenta on the Fermi surface moving along straight
classical trajectories, the direction of which is given by the
corresponding Fermi velocity. A ferromagnetic metal has different
Fermi surfaces and, correspondingly, different sets of trajectories
for the two spin orientations. In this case, the quasiclassical theory
can be used to describe two limiting cases: {\it i)} weak
ferromagnetism, where the energy splitting of the two Fermi surfaces
and the associated deviation of the Fermi velocities is so small that
the two spin trajectories with the same momentum direction are fully
coherent, and {\it ii)} strong ferromagnetism, where the splitting is
so large that the coherence is lost completely. While the former
limiting case has been exhaustively studied in the literature, the
latter has only recently received attention.\cite{PRL} Here we
present a theoretical study of the latter possibility. Even in the
absence of conventional Andreev reflection processes (which would
require coherence between particles and holes in opposite spin bands),
interesting and nontrivial physics emerges due to spin-active
interfacial scattering.  Additional motivation has been provided by
the growing interest in a new class of materials, half-metallic
ferromagnets.\cite{Pickett,Park,Soulen,Ji} In these materials, one
spin band is metallic and the other one insulating (100\% polarized
ferromagnet).  Since a half metal has a Fermi surface only for one of
the two spin orientations, it is clear that the traditional
description for weak ferromagnets is inapplicable, and other methods
must be employed.

In Sec.~II, we outline the central equations of the quasiclassical
theory of superconductivity.  Compared with the full microscopic
theory, the quasiclassical theory offers considerable simplifications
when treating spatially inhomogeneous states by reducing the content
of (unnecessary) information carried by the Green's functions.
However, this leads to nontrivial boundary conditions which have to be
formulated at interfaces separating different metals that connect the
solutions of the two sides. Such conditions have been derived for
nonmagnetic interfaces by Zaitsev,\cite{Zaitsev} and for magnetic
interfaces by Millis {\it et al.}\cite{Millis}  After a short
description of this work in Sec.~III, we formulate an alternative but
equivalent set of boundary conditions, where the transmission through
an interface is parameterized by a hopping amplitude that contains the
information of various processes contributing to particle transfer.
This approach enables the formulation of boundary conditions in a
simple and appealing form. The equivalence to existing methods is
demonstrated in Sec.~IV.  As explained in Sec.~V, the advantages of
the $t$-matrix formulation are especially evident in studying
interfaces that separate two materials with a different structure of
the Green's functions and/or a different number of trajectories, such
as in the case of a superconductor/strong ferromagnet interface.
Finally, in Sec.~VI, we apply our theory to study the current through
a point contact separating a singlet superconductor from a strong
ferromagnet.

\section{Basic equations of quasiclassical theory}

The quasiclassical theory of superconductivity 
\cite{Eilenberger,Larkin} is formulated in terms
of quasiclassical Green's functions (or propagators)
$\check{g}(\ki,{\bf R},\epsilon,t)$ that depend on the spatial
coordinate ${\bf R}$ and time $t$ and describe quasiparticles with
energy $\epsilon$ (measured from the chemical potential) and the
momentum direction on the Fermi surface $\ki={\bf p}/p_f$ moving along
classical trajectories with direction given by the Fermi velocity
${\bf v}_f(\ki)$.\cite{Serene} The quasiclassical Green's functions
are 2$\times$2 matrices in Keldysh space (denoted by a ``check''
accent),

\begin{equation}
\check g= \left(
\begin{array}{cc}
\hat g^R & \hat g^K \\
0 & \hat g^A
\end{array}
\right).
\label{Keldysh}
\end{equation}
with three nonzero elements (retarded $\hat g^R$, advanced $\hat g^A$,
and Keldysh $\hat g^K$). In describing superconductivity, these
elements in turn are 4$\times$4 Nambu-Gor'kov matrices in combined
particle-hole and spin space (denoted by the hat symbol), for example,
the retarded Green's function has the form

\begin{equation}
\hat g^R= \left(
\begin{array}{cccc}
g_{\uparrow\uparrow}^R &g_{\uparrow\downarrow}^R& 
f_{\uparrow\uparrow}^R &f_{\uparrow\downarrow}^R\\
& & & \\
g_{\downarrow\uparrow}^R &g_{\downarrow\downarrow}^R& 
f_{\downarrow\uparrow}^R &f_{\downarrow\downarrow}^R\\
& & & \\
\tilde f_{\uparrow\uparrow}^R &\tilde f_{\uparrow\downarrow}^R& 
\tilde g_{\uparrow\uparrow}^R &\tilde g_{\uparrow\downarrow}^R\\
& & & \\
\tilde f_{\downarrow\uparrow}^R &\tilde f_{\downarrow\downarrow}^R& 
\tilde g_{\downarrow\uparrow}^R &\tilde g_{\downarrow\downarrow}^R
\end{array}
\right).
\label{Nambu}
\end{equation}
All these matrix elements are not independent of each other. Indeed,
the elements in the lower half of the matrix are related to the ones
in the upper half through the conjugation symmetry, {\it e.g.}

\begin{equation}
\tilde g_{\alpha\beta}^{R,A,K}(\ki,{\bf R},\epsilon,t)
=g_{\alpha\beta}^{R,A,K}(-\ki,{\bf R},-\epsilon,t)^*.
\end{equation}
The quasiclassical Green's functions satisfy the Eilenberger transport
equation

\begin{equation}
\left[\epsilon \check \tau_3 - \check \Sigma,
\check g\right]_{\otimes} +
i \vR(\ki) \cdot \nabla_{\bf R} 
\check g=0. 
\label{eilen}
\end{equation}
Generally speaking, the self energy $\check \Sigma(\ki,{\bf
  R},\epsilon,t)$ includes molecular fields, the superconducting order
parameter $\check \Delta=\hat \Delta \check 1$, impurity scattering,
and external fields. The noncommutative product $\otimes$ combines
matrix multiplication with a convolution over the internal variables,
and $\check \tau_3=\hat \tau_3 \check 1$ is a Pauli matrix in
particle-hole space.  The quasiclassical Green's functions also
satisfy a normalization condition

\begin{equation}
 \check g \otimes \check g = -\pi^2 \check 1.
\label{normalize}
\end{equation}
In addition to (\ref{eilen}) and (\ref{normalize}), self-consistency
equations for different parts of the self-energy have to be provided;
{\it e.g.} for the (weak-coupling) order parameter the condition reads

\begin{equation}
\hat \Delta ({\bf R},t) = \lambda \int^{\epsilon_c}_{-\epsilon_c} 
\frac{d\epsilon }{4\pi i} 
\langle \hat f^{K}(\ki,{\bf R},\epsilon,t) \rangle_{\ki},
\end{equation}
where $\lambda$ is the strength of the pairing interaction, $\langle
\hspace{2mm} \rangle_{\ki}$ denotes averaging over the Fermi surface,
and $\hat f^{K}$ is the particle-hole off-diagonal part of the
quasiclassical Keldysh Green's function. The cut-off energy
$\epsilon_c$ is to be eliminated in favor of the transition
temperature in the usual manner. 

When the quasiclassical Green's function has been determined,
physical quantities of interest can be calculated; {\it e.g.}
the expression for the current density adopts the form

\begin{equation}
{\bf j}({\bf R},t) = \int
\frac{d\epsilon }{8\pi i} 
{\rm Tr}
\langle e N_f {\bf v}_f(\ki)
\hat \tau_3 \hat g^{K}(\ki,{\bf R},\epsilon,t) 
\rangle_{\ki},
\label{current}
\end{equation}
where $e$ is the electron charge and $N_f$ is the density of states on
the Fermi surface. However, to form a complete theory for studying
heterostructures, the above equations must still be supplemented with
the boundary conditions connecting the solutions at the separating
interfaces. We introduce these conditions in the following chapter.

\section{Boundary conditions}

\subsection{Scattering-matrix approach}

Interfaces represent strong perturbations on an atomic length scale
and, therefore, fall out of the applicability range of quasiclassical
theory. However, as was shown in the pioneering work of Zaitsev,
\cite{Zaitsev} interfaces can be brought within the quasiclassical
theory by means of effective boundary conditions that connect
trajectories related through interface scattering processes. Later
these conditions were generalized for an arbitrary magnetically active
interface, {\it i.e.}  one that scatters quasiparticles differently
depending on their spin orientation.\cite{Millis} The latter case is
relevant for studying interfaces with spin-polarized materials such as
ferromagnets. The procedure for the derivation of the boundary
conditions begins by isolating a region of quasiclassical size $\vert
x \vert < \delta$ around the interface located at the origin of the
perpendicular coordinate $x$ ($\delta$ much larger than the
atomic-size range of the strong interface potential but much smaller
than the superconducting coherence length $\xi$). In the half spaces
$\vert x \vert > \delta$, the solutions for quasiclassical Green's
functions can be found by standard methods described in the previous
chapter. The solutions for the left ($l$) and right ($r$) sides are
then matched via a scattering matrix

\begin{equation}
\hat {\bf S}= \left(
\begin{array}{cc}
\hat S^{ll} & \hat S^{lr}  \\
\hat S^{rl} & \hat S^{rr}
\end{array}
\right),
\label{sm}
\end{equation}
the form of which is determined by the detailed microscopic structure
of the interface region and on the quasiclassical level has to be
treated as a phenomenological parameter of the theory. The crucial
simplifying observation is that, since the strong (of the order of the
Fermi energy) interface potential dominates the Hamiltonian in the
interface region, the scattering matrix (\ref{sm}) corresponds to that
of the {\it normal state}, {\it i.e.} does not contain 
particle-hole mixing. Also, it has no Keldysh space structure.

The boundary conditions were derived for a smooth (on the scale of
$\xi$) interface, assuming the conservation of momentum ${\bf
  p}_\parallel$ parallel to the interface.  In the following, all
momentum-dependent quantities should be understood as having the same
${\bf p}_\parallel$, unless explicitly stated. In terms of
quasiclassical Green's functions they adopt the form \cite{Millis}

\newcounter{saveeqn}%
\newcommand{\alpheqn}{\setcounter{saveeqn}{\value{equation}}%
\stepcounter{saveeqn}\setcounter{equation}{0}%
\renewcommand{\theequation}
        {\mbox{\arabic{saveeqn}\alph{equation}}}}%
\newcommand{\reseteqn}{\setcounter{equation}{\value{saveeqn}}%
\renewcommand{\theequation}{\arabic{equation}}}%
\alpheqn
\begin{eqnarray}
\label{millis}
\nonumber
(\check g^l_{in}-i\pi \check 1) \otimes(\hat S_{ll}^\dagger 
\check g^l_{out}\hat S_{ll}
-\hat S_{rl}^\dagger \check g^r_{out}\hat S_{rl})
&\otimes&(\check g^l_{in}+i\pi \check 1) \\ &=&0, \\
\nonumber
(\check g^l_{out}+i\pi \check 1)\otimes (\hat S_{ll}\check 
g^l_{in}\hat S_{ll}^\dagger
-\hat S_{lr}\check g^r_{in}\hat S_{lr}^\dagger)
&\otimes&(\check g^l_{out}-i\pi \check 1) \\ &=&0, \\
\nonumber
(\check g^r_{in}-i\pi \check 1)\otimes (\hat S_{rr}^\dagger 
\check g^r_{out}\hat S_{rr}
-\hat S_{lr}^\dagger \check g^l_{out}\hat S_{lr})
&\otimes&(\check g^r_{in}+i\pi \check 1) \\ &=&0, \\ 
\nonumber
(\check g^r_{out}+i\pi \check 1)\otimes (\hat S_{rr}\check 
g^r_{in}\hat S_{rr}^\dagger
-\hat S_{rl}\check g^l_{in}\hat S_{rl}^\dagger)
&\otimes&(\check g^r_{out}-i\pi \check 1) \\ &=&0, 
\end{eqnarray}
\reseteqn with $\check g_{in}= \check g(\ki)$ and $\check
g_{out}=\check g(\ko)$, where $\ki$ ($\ko$) is a unit vector along the
momentum direction with the perpendicular component directed towards
(away from) the interface. The boundary condition consists of four
coupled nonlinear equations for the incoming and outgoing matrix
propagators on both sides of the interface. Solving this equation
system and dealing with the possibility of arriving at unphysical
solutions is evidently not a simple task. Progress towards a more
convenient form of boundary conditions has been made by Eschrig
(nonmagnetic interfaces) \cite{Eschrig} and Fogelstr\"om (magnetic
interfaces).\cite{Fogelstrom} They employed the powerful Riccati
parameterization method which allows for a considerably simpler
representation of boundary conditions in terms of the Riccati
amplitudes.\cite{Schopohl} However, the conditions in
Ref.~\onlinecite{Fogelstrom} were only derived for the equilibrium (retarded
and advanced) propagators. Furthermore, even in equilibrium
situations they can not be used in the published form in the case when
the two sides of the interface have a different number of trajectories
({\it i.e.} when matrices $\hat S_{lr}$ and $\hat S_{rl}$ are not
invertable). This situation arises in the context of half-metallic
materials, where trajectories exist only for one of the spin
orientations.

\subsection{Transfer-matrix approach}

Due to the abovementioned difficulties we proceed in an alternative
but equivalent route.\cite{Cuevas} This method requires solving for
the auxiliary quasiclassical propagators $\check g^{l,0}$ and $\check
g^{r,0}$ for an {\it impenetrable} interface. They are to be
calculated with the self-energies $\check \Sigma\{\check g\}$
determined with the full propagator, and using the simple
perfectly-reflecting boundary condition

\begin{equation}
\check g^{i,0}_{out}=\hat S^i \check g^{i,0}_{in} (\hat S^{i})^\dagger,
\end{equation}
where $i=l,r$. They also satisfy the normalization condition, $\check
g^{i,0} \otimes \check g^{i,0}=-\pi^2 \check 1$. The impenetrable
interface is characterized by two surface scattering matrices, $\hat
S^l$ and $\hat S^r$. Particle conservation requires them to be
unitary, $(\hat S^i)^\dagger=(\hat S^i)^{-1}$. The transmission
processes for an interface with arbitrary transparency can be taken
into account with a $t$-matrix formulation.  The transfer matrices are
determined with effective hopping amplitudes $\hat \tau_{lr}$ and
$\hat \tau_{rl}$ by the following equations:

\alpheqn
\begin{eqnarray}
\label{tmatrix}
\check t^l_{in}&=& 
\hat \tau_{lr} \; {\check g}^{r,0}_{out} \; 
\hat \tau_{lr}^{\dagger }
+ \hat \tau_{lr} \; {\check g}^{r,0}_{out} \; \hat \tau_{lr}^{\dagger }
\otimes {\check g}^{l,0}_{in} \otimes \check t^l_{in}, \\
\check t^r_{in}&=& 
\hat \tau_{rl} \; {\check g}^{l,0}_{out} \;
\hat \tau_{rl}^{\dagger }
+ \hat \tau_{rl} \; {\check g}^{l,0}_{out} \; \hat \tau_{rl}^{\dagger }
\otimes {\check g}^{r,0}_{in} \otimes \check t^r_{in}, 
\end{eqnarray}
\reseteqn
with $\hat \tau_{rl}=(\hat \tau_{lr})^\dagger$
due to particle conservation. The corresponding
$t$ matrices for outgoing trajectories are related to the ones
for incoming trajectories through the relation

\begin{equation}
\check t^i_{out}=\hat S^i \check t^i_{in} (\hat S^i)^\dagger.
\end{equation}
The $t$ matrix describes the modifications of the decoupled
quasiclassical propagators due to virtual hopping
processes to the opposite side. Finally, the boundary condition
can be expressed in terms of $\check t^{i}$ and $\check g^{i,0}$
to read  
\alpheqn
\begin{eqnarray}
\check g^i_{in}\!\!&=&\!\!\check g^{i,0}_{in} +
( \check g^{i,0}_{in} + i\pi\check 1 ) \otimes 
\check t^i_{in} \otimes (\check g^{i,0}_{in} - i\pi\check 1), 
\hspace{0.5cm}\\
\check g^i_{out}\!\!&=&\!\!\check g^{i,0}_{out} +
(\check g^{i,0}_{out} - i\pi\check 1) \otimes
 \check t^i_{out} \otimes (\check g^{i,0}_{out} + i\pi\check 1). 
\hspace{0.5cm}\label{eq7}
\end{eqnarray}
\reseteqn In the $t$-matrix description, the phenomenological
parameters containing the microscopic information of the interface are
the surface scattering matrices and the hopping amplitudes. The
particle-hole structures of the surface scattering matrix and the
hopping amplitude are connected through

\begin{equation}
\hat S^i=\left(
\begin{array}{cc}
S^{i} & 0 \\
0 & \tilde S^{i}
\end{array}
\right), \; \;
\hat \tau_{lr}=\left(
\begin{array}{cc}
\tau_{lr} & 0 \\
0 & (\tilde S^{l})^\dagger \tau_{lr}^* (\tilde S^{r})^\dagger
\end{array}
\right),
\end{equation}
to ensure the conservation of current. In the general case

\begin{equation} 
\tilde S ({\bf p}_\parallel)=S^{tr}(-{\bf p}_\parallel).
\label{tilde}
\end{equation}
In this formulation, the boundary problem effectively reduces to
calculating the auxiliary Green's functions for perfectly reflecting
interfaces. Numerically this is an extremely simple task, {\it e.g.},
employing the procedure of Riccati parameterization. Afterwards the
boundary Green's functions for the partially transmitting interface
can be obtained directly from Eqs.~(13), since solving for the
necessary $t$ matrices (11) only involves a 4$\times$4 matrix
inversion.  When contrasted with solving the group of equations (9),
the $t$-matrix approach manifests its usefulness.

\section{Relation to other methods}

The underlying perturbative nature of the $t$-matrix approach might
arise suspicions concerning its applicability when the interface in
question has high transparency. The boundary conditions (13) are,
however, valid for arbitrary transmission and, in fact, completely
equivalent to the corresponding scattering-matrix description (9).
The connection between the two approaches is established by the
following identification of the full scattering matrix in terms of the
surface scattering matrices and hopping amplitudes:

\begin{equation}
\label{scatt}
\hat {\bf S}= 
\left(
\begin{array}{cc}
\hat S_{ll} & \hat S_{lr} \\ \hat S_{rl} & \hat S_{rr}
\end{array}
\right)
=
\left(
\begin{array}{cc}
\hat S_l & 0 \\ 0 & {\hat 1}
\end{array}
\right)
\left(
\begin{array}{cc}
\hat r & \hat d \\
\hat d^{\dagger} & 
-\underline{\hat r} 
\end{array}
\right)
\left(
\begin{array}{cc}
\hat 1 & 0 \\ 0 & \hat S_r
\end{array}
\right),
\end{equation}
where we have defined

\alpheqn
\begin{eqnarray}
\hat r &=& (1+\pi^2\hat\tau_{lr} \hat\tau_{rl})^{-1}
(1-\pi^2\hat\tau_{lr} \hat\tau_{rl}), \\
\underline{\hat r} &=& (1+\pi^2\hat\tau_{rl}\hat\tau_{lr} )^{-1}
(1-\pi^2\hat\tau_{rl}\hat\tau_{lr} ), \; \; {\rm and} \\
\hat d &=& (1+\pi^2\hat\tau_{lr} \hat\tau_{rl})^{-1} 
2\pi \hat\tau_{lr}. 
\end{eqnarray}
\reseteqn The identity (\ref{scatt}) serves as a precise definition
of the auxiliary matrices $\hat S_l$ and $\hat S_r$ in terms of the
physical parameters of the full scattering matrix.  Using
(\ref{tilde}), the particle ($S_p$) and hole ($S_h$) parts of
(\ref{scatt}) can be seen to be related by

\begin{equation}
\label{scattph}
S_h({\bf p}_\parallel)=
\left( \begin{array}{cc} \tilde S_l & 0 \\ 0 & \tilde S_r \end{array}
\right)
S^*_p(-{\bf p}_\parallel)
\left( \begin{array}{cc} \tilde S_l & 0 \\ 0 & \tilde S_r \end{array}
\right).
\end{equation}
In particular, if the interface scattering matrix is spin-inactive,
Eqs.~(9) reduce to those derived by Zaitsev. In the following, we
show that the solution of (13) in the appropriate limit also solve
Zaitsev's boundary conditions for arbitrary transmission of the
interface, and in both equilibrium and nonequilibrium situations.
On the other hand, in the case of diffusive conductors the
boundary conditions of the $t$-matrix approach are equivalent to the
ones derived by Nazarov.\cite{Nazarov}

\subsection{Zaitsev's boundary conditions}

The boundary conditions of Zaitsev read (we suppress the symbol
$\otimes$ and unit matrices for clarity) \cite{Zaitsev}

\alpheqn
\begin{eqnarray}
\label{z1}
\check g^l_a=\check g^r_{a}&=&\check g_a, \\
\check g_a[R(\check g_s^{+})^2+(\check g_s^{-})^2]&=&
-i\pi D\check g_s^{-} \check g_s^{+},
\label{z2}
\end{eqnarray}
\reseteqn where $R$ ($D$) is the reflection (transmission)
coefficient, $R+D=1$, $\check g^{l,r}_a=\pm(\check g^{l,r}_{in}-\check
g^{l,r}_{out})/2$, and $\check g_s^{+,-}=(\check g_s^r \pm \check
g_s^l)/2$, with $\check g_s^{l,r}=(\check g^{l,r}_{in}+\check
g^{l,r}_{out})/2$. In the corresponding limiting case the surface
scattering matrices $\hat S^{l,r}$ are unit matrices, the hopping
element can be taken as a real number, $\hat \tau_{lr}=\tau \hat 1$,
and the boundary conditions in the $t$-matrix approach are

\alpheqn
\begin{eqnarray}
\check g^i_{in}&=&\check g^{i,0}+
( \check g^{0} + i\pi ) 
\; \check t^i \; (\check g^{i,0} - i\pi ), 
\\
\check g^i_{out}&=&\check g^{i,0}+
(\check g^{i,0} - i\pi ) 
\; \check t^i \; (\check g^{i,0} + i\pi ),
\label{ours}
\end{eqnarray}
\reseteqn
with $\check g^{i,0}_{in}=\check g^{i,0}_{out}=\check g^{i,0}$
and $\check t^i_{in}=\check t^i_{out}=\check t^i$. The $t$-matrix
equations now take the form

\alpheqn
\begin{eqnarray}
\label{tours}
\check t^l&=& (1-\tau^2 {\check g}^{r,0} {\check g}^{l,0})^{-1}
\tau^2 {\check g}^{r,0}, \\
\check t^r&=& (1-\tau^2 {\check g}^{l,0} {\check g}^{r,0})^{-1}
\tau^2 {\check g}^{l,0}.
\end{eqnarray}
\reseteqn
From (17) we have

\begin{equation}
\check g^l_a=i\pi [\check t^l,\check g^{l,0}], \hspace{5mm}
\check g^r_a=-i\pi [\check t^r,\check g^{r,0}],
\label{anom}
\end{equation}
which, using (18) and the identity

\begin{equation}
(1-\check a \check b)^{-1} \check a = \check a (1-\check b \check a)^{-1},
\label{ide}
\end{equation}
immediately gives (\ref{z1}). This condition ensures the conservation
of current. To show (\ref{z2}), we first express it in terms of
the quantities $\check g_s^{l,r}$ as follows:

\begin{eqnarray}
\nonumber
(1-R)\left[ \left(1+\frac{\check g_a}{i\pi} \right)\check g_s^l 
\check g_s^r - \left(1-\frac{\check g_a}{i\pi} \right)\check g_s^r 
\check g_s^l
\right] \\ - 2i\pi (R+1)\check g_a \left[ 1-
\left(\frac{\check g_a}{i\pi} \right)^2
\right]=0,
\label{equ1}
\end{eqnarray} 
where we have used the identity

\begin{equation}
(\check g_s^i)^2+(\check g_a)^2=-\pi^2,
\end{equation}
$i=l,r$. Using (17) we find

\begin{eqnarray}
\nonumber
\check g_s^l \check g_s^r &=& \left(1-\frac{\check g_a}{i\pi} \right)^2
\check g_s^{l,0} \check g_s^{r,0}, \\
\check g_s^r \check g_s^l &=& \left(1+\frac{\check g_a}{i\pi} \right)^2
\check g_s^{r,0} \check g_s^{l,0},
\end{eqnarray}
whereby Eq.~(\ref{equ1}) transforms to

\begin{eqnarray}
\nonumber
\left[ 1-\left(\frac{\check g_a}{i\pi} \right)^2 \right]
\Big\{ (1-R)\Big[ 
\left(1-\frac{\check g_a}{i\pi} \right)\check g_s^{l,0} 
\check g_s^{r,0} - 
\\ \left(1+\frac{\check g_a}{i\pi} \right)
\check g_s^{r,0} \check g_s^{l,0}
\Big]   
- 2i\pi (R+1)\check g_a \Big\}
=0.
\label{equ2}
\end{eqnarray} 
This form exhibits directly the unphysical solutions of Zaitsev's
boundary conditions, determined by vanishing of
the first square bracket in (\ref{equ2}).
The physical solutions are given by the requirement that the
curly bracket of (\ref{equ2}) vanishes. On inserting (18) and (19)
into this expression and using the identity

\begin{equation}
(1-\check a)^{-1} \check a - (1-\check b)^{-1} \check b
= (1-\check a)^{-1} - (1-\check b)^{-1},
\end{equation}
one arrives at the condition

\begin{equation}
\left[ (1-R)(1+\pi^4 \tau^4)-2(R+1)\pi^2 \tau^2 \right] \check g_a=0,
\end{equation}
which is identically fulfilled provided that the transmission
coefficient in the $t$-matrix description is identified as

\begin{equation}
D=1-R=\frac{4\pi^2 \tau^2}{(1+\pi^2 \tau^2)^2}.
\label{transm}
\end{equation}

\subsection{Nazarov's boundary conditions}

The boundary conditions for diffusive conductors, presented by Nazarov,
\cite{Nazarov} are formulated in terms of a Keldysh-Nambu matrix
current, the Keldysh part of which defines the electric current
through the interface. In the $t$-matrix approach, this matrix is
proportional to $\check g_a$ of Eq.~(\ref{anom}) and, therefore, to
the quantity

\begin{equation}
\check I = [ \check t^l,\check g^{l,0}],
\end{equation}
determined at the left-hand side of the interface. To simplify the
following expressions, we again choose $\hat \tau_{lr}=\tau \hat 1$
and real. Furthermore, in the context of diffusive conductors both the
Green's functions and the hopping elements should be regarded
as trajectory-averaged quantities, {\it i.e.} independent of $\ki$.
Using (21a) and (\ref{ide}) we obtain

\begin{equation}
\check I = \tau^2 \check g^{r} \check g^{l}
(\check 1 - \tau^2 \check g^{r} \check g^{l})^{-1}
- \tau^2 \check g^{l} \check g^{r}
(\check 1 - \tau^2 \check g^{l} \check g^{r})^{-1},
\end{equation}
where we have dropped the zero from the superscript (all Green's
functions are auxiliary ones). Writing the matrix current in the form

\begin{eqnarray}
\nonumber
\check I = \tau^2 \check g^{r} \check g^{l}
(\check 1 - \tau^2 \check g^{l} \check g^{r})
(\check 1 - \tau^2 \check g^{l} \check g^{r})^{-1}
(\check 1 - \tau^2 \check g^{r} \check g^{l})^{-1} \\
-\tau^2 \check g^{l} \check g^{r}
(\check 1 - \tau^2 \check g^{l} \check g^{r})^{-1}
(\check 1 - \tau^2 \check g^{r} \check g^{l})
(\check 1 - \tau^2 \check g^{r} \check g^{l})^{-1}, \hspace{2mm}
\end{eqnarray}
and exploiting the fact that $\check g^{l} \check g^{r}$ commutes
with $\check g^{r} \check g^{l}$, we arrive at

\begin{eqnarray}
\nonumber
\check I &=& -\tau^2 [\check g^{l}, \check g^{r}]
(\check 1 - \tau^2 \check g^{l} \check g^{r})^{-1}
(\check 1 - \tau^2 \check g^{r} \check g^{l})^{-1} \\
&=& -\tau^2 [\check g^{l}, \check g^{r}]
(1-\tau^2 \{\check g^{l}, \check g^{r}\}+\pi^4 \tau^4)^{-1}.
\end{eqnarray}
Finally, using Eq.~(\ref{transm}) to identify the transmission
coefficient, and defining $\check G^{l,r}\equiv\check g^{l,r} /(i\pi)$
because of the different convention for normalizing the
Green's functions used in Ref.~\onlinecite{Nazarov},
we arrive at

\begin{equation}
\check I = 
\frac{D [\check G^{l}, \check G^{r} ]}
{4+D
(\{ \check G^{l},\check G^{r} \} 
-2)},
\end{equation}
which, apart from the prefactor, is the matrix-current expression
defining the boundary conditions of Nazarov.

\section{Interface problem with ferromagnets}

\subsection{Weak and strong ferromagnetism}

\begin{figure}[bt]
\begin{center}
\epsfxsize=0.5\textwidth{\epsfbox{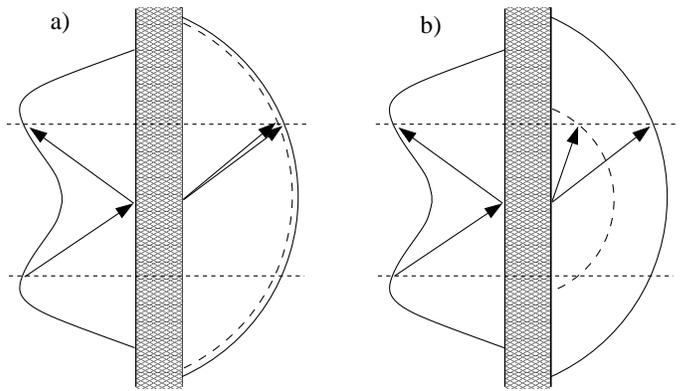}}
\caption[interface]{
  Two quasiclassical pictures of an interface separating a
  spin-unpolarized material (left-hand side of the interface) from a
  ferromagnet (right-hand side): a) weak ferromagnet with a small
  splitting of Fermi surfaces for the two spin orientations (indicated
  by solid and dashed curves) and b) strong ferromagnet with a large
  splitting. Incoming trajectory from the left-hand side and
  corresponding outgoing ones with the same parallel momentum on both
  sides are indicated by arrows.  }
\label{interface}
\end{center}
\end{figure}

As already mentioned, the quasiclassical theory is formulated in terms
of quasiparticles travelling along classical trajectories. Smooth
interfaces between different materials introduce coupling between
incoming and outgoing trajectories with the same momentum parallel to
the interface. A ferromagnet has a different Fermi surface (or,
equivalently, set of trajectories) for each of the two possible spin
orientations. Consequently, two different limiting cases that allow a
quasiclassical description naturally emerge (see
Fig.~\ref{interface}). In the first case the exchange energy splitting
of the two Fermi surfaces is small enough that the quasiparticle wave
packets on the two trajectories corresponding to the same parallel
momentum but different spins overlap and, therefore, the two
trajectories remain fully coherent in the ferromagnetic region
(Fig.~\ref{interface}a). Technically this means that the full
2$\times$2 spin structure of the quasiclassical Green's functions,
defined by Eq.~(\ref{Nambu}), is to be retained in the ferromagnetic
side of the interface. This case, relevant for weak ferromagnets, has
been widely studied in the literature; the standard description simply
involves a spin-dependent shift in the quasiparticle energy, effected
by the replacement

\begin{equation}
\epsilon \hat \tau_3
\rightarrow \epsilon \hat \tau_3 -h\sigma_3 \hat 1
\label{shift}
\end{equation}
in the Eilenberger equation (\ref{eilen}). Here $h$ is the
exchange-field parameter and $\sigma_3$ is a Pauli spin matrix.
Other Fermi-surface parameters, {\it i.e.} Fermi velocities and
the density of states, are assumed identical for the two spin
bands in the ferromagnet. 

In this article we restrict ourselves to the opposite limiting case of
strongly ferromagnetic materials, illustrated in
Fig.~\ref{interface}b. That is, we assume the exchange splitting and
the resulting directional deviation of the two spin trajectories
sharing the same parallel momentum to be so large that the coherence
between them is lost completely. As a consequence, the quasiclassical
propagators have no matrix structure in spin space. In particular,
conventional Andreev reflection processes are forbidden because
electrons and holes in opposite spin bands occupy different
trajectories which do not interfere with each other. Trajectories with
different spin orientations can only be coupled incoherently, such as
{\it e.g.} due to elastic spin-flip scattering by magnetic impurities.
It should be emphasized that no energy shift of the form (\ref{shift})
should be introduced in this limit; instead, Fermi velocities and the
density of states become spin dependent. The reason for this is that
the integration over the energy of relative motion (``$\xi$
integration''), employed in the formal process of converting the full
two-particle Green's function into quasiclassical ones, is now
performed separately around the two different Fermi surfaces. This is
in contrast to the case of weak ferromagnets, where the same
$\xi$-integration range is used for both Fermi surfaces
simultaneously.

A very interesting special case which falls into the latter category
of ferromagnets with strong spin splitting is that of half-metallic
materials. In fact, half metals are metallic in one of the spin bands
only -- the other one is insulating. Such behaviour has recently been
reported in CrO$_2$ \cite{Soulen,Ji} and in certain manganite
materials.\cite{Park} and has attracted considerable attention
because of possible applications in the emerging field of spintronics.
\cite{Wolf} Since in half metals a Fermi surface only exists for one
of the spin orientations, the standard description for weak
ferromagnets is obviously inapplicable. However, half metals still
allow for a straightforward quasiclassical treatment in the
separate-band picture: quasiparticle trajectories simply exist only
for one of the spin orientations.

\subsection{Spin mixing}

The quasiclassical boundary conditions in the hopping description
involve surface scattering matrices $\hat S^{l,r}$ that characterize a
fully reflecting interface. In the case of a magnetically
active interface the most general form of such matrices (for
quasiparticles), satisfying the requirement of unitarity, was pointed
out by Tokuyasu {\it et al.} \cite{Tokuyasu} to be

\begin{equation}
S=e^{-i\Phi/2}e^{-i(\theta/2)\hat{\mbox{\boldmath $\mu$}} \cdot 
\mbox{\boldmath $\sigma$}},
\end{equation}
where $\hat{\mbox{\boldmath $\mu$}}$ is a unit vector pointing to the
direction of the surface magnetization and $\mbox{\boldmath $\sigma$}$
is a vector constructed of Pauli spin matrices. The corresponding
scattering matrix for quasiholes follows from Eq.~(\ref{tilde}).
Dropping the irrelevant overall phase factor $\Phi$, the surface
scattering matrix is determined by a single parameter, the spin-mixing
angle $\theta$. The physics behind spin mixing can be visualized as
follows: even for a fully reflecting interface, incident wave
functions penetrate a small distance into the forbidden,
spin-polarized region. This results in different matching conditions
for waves with opposite spin directions and, consequently, different
phase shifts for the reflected waves.

The relative phase difference introduced by spin mixing results in
interesting nontrivial phenomena at superconductor/ferromagnet
interfaces, even in the absence of quantum-mechanical coherence
between the two spin bands in the ferromagnet. One such example is the
recent prediction of a nonvanishing Josephson current in a
heterostructure with a mesoscopic half-metallic piece separating two
singlet superconductors -- driven by spin-triplet 
pairing correlations.\cite{PRL}
(this effect requires, in addition to spin mixing, also the presence
of spin-flip centers at the interfaces). However, even though spin
mixing is expected to be an intrinsic feature of any spin-active
interface, systematic experimental estimations of the typical
magnitudes of $\theta$ are not yet available. As a guideline for such
future experiments, we study in the following chapter the differential
conductance of a spin-mixing point contact between a singlet
superconductor and a strong ferromagnet -- simultaneously offering a
view of the $t$-matrix approach at work. 

\section{S/F point contact with spin mixing}

We consider a point contact with arbitrary transmission separating a
conventional singlet superconductor and a strong ferromagnetic
material. The small (compared with the coherence length of the
superconductor) dimensions of the contact and, consequently, the small
size of the current flowing through it does not appreciably affect the
state of the coupled half-spaces from that corresponding to
zero transmission.  This
offers a simplification by relieving us from the necessity of
calculating the superconducting order parameter self-consistently.
According to Eq.~(\ref{current}), the current, calculated at the
interface on the ferromagnetic side, adopts the form

\begin{equation}
j=\sum_\alpha \int \frac{d\epsilon}{8\pi i}
\langle e N_f^\alpha v_f^\alpha \cos\phi \hspace{1mm}
{\rm Tr} [\hat \tau_3 (\hat g_{in}^K - \hat g_{out}^K)] 
\hspace{1mm} \rangle^\alpha_+,
\label{curr1}
\end{equation}
where $\alpha=\uparrow,\downarrow$ labels the spin band of the
ferromagnet, each with its own density of states $N_f^\alpha$ and the
Fermi velocity $v_f^\alpha$. For simplicity, the Fermi surfaces are
assumed cylindrical and the interface specularly reflecting, the
generalizations are straightforward. The impact angle $\phi$ determines
the angle between the trajectory and the current direction. The
angular averaging is to be taken over trajectories with $\cos\phi \geq
0$. The two spin bands in the ferromagnet give two separate
contributions to the current. From Eqs.~(13) follows

\begin{equation}
\hat g_{in}^K - \hat g_{out}^K = 2\pi i \; [\check t,\check g^{0}]^K,
\label{commu}
\end{equation}
where the $t$ matrix and the auxiliary Green's function $\check g^{0}$
(for a perfectly reflecting interface) are to be evaluated on the
ferromagnetic side where the latter has the simple form $\hat
g^{R,0}=-\hat g^{A,0}=-i\pi\hat \tau_3$, and $\hat g^{K,0}=\hat
g^{R,0}\hat F-\hat F\hat g^{A,0}$, with

\begin{equation}
\hat F \equiv \left(
\begin{array}{cc}
F_e  & 0  \\
0 & F_h
\end{array}
\right)= 
\left(
\begin{array}{cc}
\tanh\left( \frac{\epsilon-eV}{2T} \right)  & 0  \\
0 & \tanh\left( \frac{\epsilon+eV}{2T}\right)
\end{array}
\right),
\end{equation}
where $V$ is the voltage over the contact and $T$ is the temperature.
We choose the electrical potential to be zero on the superconducting
side of the interface. Writing out the commutator (\ref{commu}),
Eq.~(\ref{curr1}) reads

\begin{equation}
j=\sum_\alpha {i \pi e N_f^\alpha v_f^\alpha \over 2}
\int d\epsilon
\langle \cos\phi \hspace{1mm}
{\rm Tr} [\hat t^K - (\hat t^R \hat F - \hat F \hat t^A)] 
\hspace{1mm} \rangle^\alpha_+.
\end{equation}
Using now Eq.~(11), the relation
$\hat t^A = \hat \tau_3 (\hat t^R)^\dagger \hat \tau_3$,
and the properties of the trace, we find

\begin{eqnarray}
\nonumber
j=\sum_\alpha \pi e N_f^\alpha v_f^\alpha
\int d\epsilon \; {\rm Im}
\langle \cos\phi \hspace{2cm} \\ \hspace{1mm}
{\rm Tr} [(\hat N^R)^\dagger \hat v^R \hat N^R (\hat F - F_0)] 
\hspace{1mm} \rangle^\alpha_+,
\label{curr2}
\end{eqnarray}
where we have defined an effective interface potential $\hat v^R =
\hat \tau \hat g^S \hat \tau^\dagger$, with $\hat g^S=\hat
g^{R,0}_{out}$ the auxiliary Green's function on the
superconducting side of the interface, 
$\hat N^R = (1+i\pi \hat \tau_3 \hat v^R)^{-1}$, and
$F_0=\tanh(\epsilon/2T)$. We assume that the interface does not flip
the spin, {\it i.e.} hopping processes from an incoming trajectory in
the ferromagnet to an outgoing trajectory on the superconducting side
are without loss of generality determined by two real numbers, $\hat
\tau_\alpha = \tau_\alpha \hat 1$, for the two possible spin
orientations. In this case Eq.~(\ref{curr2}) gives

\begin{equation}
j=\sum_\alpha e N_f^\alpha v_f^\alpha
\int d\epsilon \; 
\Big\langle \cos\phi
\frac{2\pi \tau_\alpha^2 \; {\rm Im} g^S_{\alpha\alpha}}
{\vert 1+i\pi \tau_\alpha^2 \; g^S_{\alpha\alpha} \vert^2}
\Big\rangle^\alpha_+ (F_e - F_0),
\label{curr3}
\end{equation}
where $g^S_{\uparrow\uparrow}$ ($g^S_{\downarrow\downarrow}$) is the
1,1 (2,2) element of the full 4$\times$4 auxiliary Green's function at
the interface on the superconducting side. In the presence of
spin mixing (described by the spin-mixing angle $\theta$) 
this has the form

\begin{equation}
g^S_{\uparrow\uparrow}=\pi \; \frac{\epsilon \cos\frac{\theta}{2}
+\Omega \sin\frac{\theta}{2}}{\epsilon \sin\frac{\theta}{2}
-\Omega \cos\frac{\theta}{2}},
\label{upup}
\end{equation}
where $\Omega \equiv \sqrt{\Delta^2-\epsilon^2}$, 
$\Delta$ 
is the magnitude of the bulk order parameter, 
and 
$g^S_{\downarrow\downarrow}$ can be obtained by replacing $\theta \rightarrow
- \theta$. Inserting (\ref{upup}) into (\ref{curr3}) we obtain

\begin{equation}
j=\sum_\alpha
e N_f^\alpha v_f^\alpha \int d\epsilon \;
\langle \; \cos\phi \; j_\epsilon^\alpha \; \rangle^\alpha_+
\; (F_e - F_0),
\label{curr5}
\end{equation}
with 

\begin{equation}
j^\uparrow_\epsilon=
\frac{2\pi^2 \tau_\uparrow^2 \; \epsilon \; {\rm Im} \; \Omega}
{\vert 
\epsilon (\sin\frac{\theta}{2}+i\pi^2 \tau_\uparrow^2 \cos\frac{\theta}{2})
-\Omega (\cos\frac{\theta}{2}-i\pi^2 \tau_\uparrow^2 \sin\frac{\theta}{2})
\vert^2}, \\
\label{curr4}
\end{equation}
and $j^\downarrow_\epsilon$ follows from $\tau_\uparrow \rightarrow
\tau_\downarrow$ and $\theta \rightarrow - \theta$.  For subgap
energies, $\vert \epsilon \vert \leq \Delta$, $j^\alpha_\epsilon$
vanishes because $\Omega$ is real. This simply reflects the fact that
the contribution from Andreev reflection processes vanishes in
quasiclassical approximation due to the lack of coherence between
spin-up and spin-down bands on the ferromagnetic side. Introducing the
normal-state transmission and reflection coefficients with
Eq.~(\ref{transm}), Eq.~(\ref{curr4}) can be written for $\vert
\epsilon \vert \geq \Delta$ as

\begin{eqnarray}
\nonumber
j^\alpha_\epsilon=\hspace{6cm}\\
 \frac{-2D^\alpha \sqrt{1-\left(\frac{
\Delta}{\epsilon}\right)^2}}
{
\left[1\!-\!\sqrt{R^\alpha}\!+\!(1\!+\!\sqrt{R^\alpha})
\sqrt{1\!-\left(\frac{\Delta}
{\epsilon}\right)^2}\right]^2 \!\!\!\!
+4\sqrt{R^\alpha}\left(\frac{\Delta}{\epsilon}\right)^2\!\sin^2\frac{\theta}{2}
}.\nonumber \\ 
\end{eqnarray}
The differential conductance $G=\partial j/\partial V$ for $\vert eV
\vert \geq \Delta$ can now be obtained by differentiation, and at
$T=0$ adopts the form

\begin{equation}
G=-\sum_\alpha
2e^2 N_f^\alpha v_f^\alpha \;
\langle \; \cos\phi \; j_\epsilon^\alpha(\epsilon=eV)
 \; \rangle^\alpha_+ \;.
\end{equation}
In particular, for a half metal with a conducting spin-up band
and a reflection coefficient $R^\uparrow=R$
independent of impact angle $\phi$, the conductance (normalized to the
normal-state value $G_N$) reads

\begin{figure}[bt]
\begin{center}
\epsfxsize=0.48\textwidth{\epsfbox{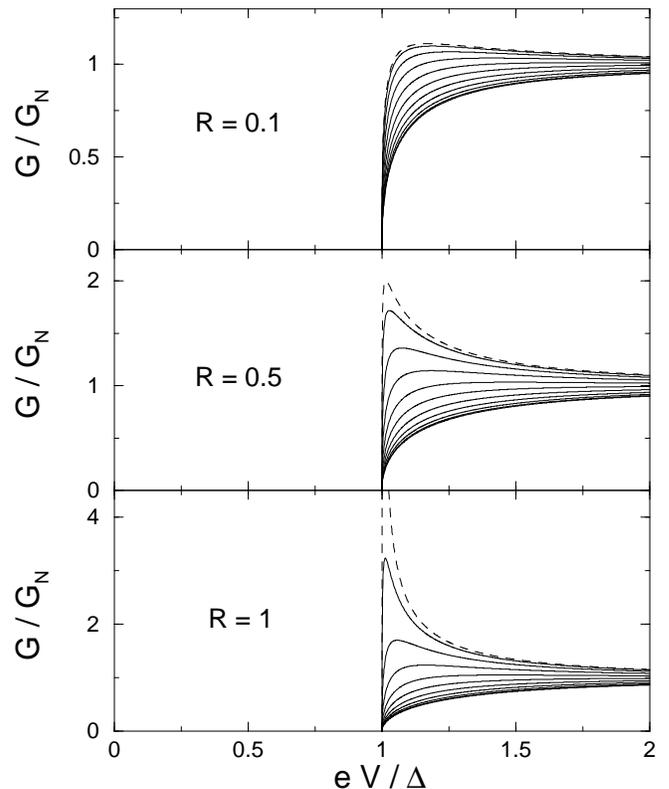}}
\caption[interface]{
  The normalized conductance $G/G_N$ as a function of $eV/\Delta$ for
  $R=0.1$ (top figure), $R=0.5$ (middle), and $R=1$ (bottom).  The
  different curves in each figure correspond, from top to bottom, to
  different values of the spin-mixing angle ranging from $\theta=0$
  (dashed curve) to $\theta=\pi$ in intervals of $\pi/10$.  }
\label{kuva05}
\end{center}
\end{figure}
\begin{eqnarray}
\nonumber
\frac{G}{G_N}=\hspace{6cm}\\
 \frac{4\sqrt{1-\left(\frac{
\Delta}{eV}\right)^2}}
{
\left[1\!-\!\sqrt{R}+(1\!+\!\sqrt{R})\sqrt{1\!-\left(\frac{\Delta}
{eV}\right)^2}\right]^2\!\!\!\!
+4\sqrt{R}\left(\frac{\Delta}{eV}\right)^2 \sin^2\frac{\theta}{2}
}, \nonumber \\
\label{conduc}
\end{eqnarray}
when $\vert eV \vert \geq \Delta$. The contribution due to a finite
spin-mixing angle $\theta$ has the effect of broadening the
conductance features near the gap edge. This is demonstrated in
Fig.~\ref{kuva05} which shows the normalized conductance as a function
of the spin-mixing angle for three different reflection coefficients
of the contact. In particular, the characteristic BCS square-root
singularity for a tunnel-limit ($R \rightarrow 1$) contact is removed.
On the other hand, for perfectly transmitting interfaces, $R
\rightarrow 0$, spin mixing has no effect. As an additional detail,
the maximum of Eq.~(\ref{conduc}), attained at
$eV/\Delta=(1+\sqrt{R})/2R^{1/4}$ when $\theta=0$, is shifted towards
higher voltages when $\theta > 0$, vanishing altogether if $\theta
\geq \pi/2$.

\section{Conclusions}

We have presented a quasiclassical theory which is suited for detailed
studies of heterostructures consisting of a wide variety of materials:
superconductors (both conventional and unconventional), normal metals,
and both weak and strong ferromagnets. The most crucial part of this
description is the treatment of boundary conditions at interfaces
separating different materials. These conditions are formulated in
terms of hopping amplitudes, containing the information of allowed
transmission processes, and the corresponding $t$ matrices. Compared
with the traditional scattering-matrix approach, the $t$ matrix
approach provides clear advantages for studying spin-active
interfaces, or interfaces which connect materials with different
numbers of trajectories or with different internal structures of their
Green's functions. A particular example are strong ferromagnets of
which the half-metallic materials form a special case. In connection
with such materials, nontrivial physics arises due to spin-dependent
interfacial scattering processes. The crucial parameter controlling
the details of these effects is the degree of spin mixing. At present,
there have been no attempts to determine experimentally the magnitude
of this parameter at a spin-active interface. To provide a guideline
for such studies, and to demonstrate the $t$-matrix approach, we have
calculated the differential conductance for a superconductor/half
metal point contact. In the tunneling limit of such contacts, the
conductance depends strongly on spin mixing, and should provide an
effective means of determining the importance of the new physics
related to spin-active interfaces.

\end{document}